\documentclass[12pt]{iopart}
\usepackage{graphicx}
\usepackage{siunitx}

\begin{document}

\title[Application of iterative phase-retrieval algorithms to ARPES orbital tomography]{Application of iterative phase-retrieval algorithms to ARPES orbital tomography}

\author{P Kliuiev, T Latychevskaia, J Osterwalder, M Hengsberger and L Castiglioni}

\address{Department of Physics, University of Zurich, Winterthurerstrasse 190, 8057 Zurich, Switzerland}
\ead{kliuiev@physik.uzh.ch, luca.castiglioni@physik.uzh.ch}
\vspace{10pt}
\begin{indented}
\item[] June 2016
\end{indented}

\begin{abstract}
Electronic wave functions of planar molecules can be reconstructed via inverse Fourier transform of angle-resolved photoelectron spectroscopy (ARPES) data, provided the phase of the electron wave in the detector plane is known. Since the recorded intensity is proportional to the absolute square of the Fourier transform of the initial state wave function, information about the phase distribution is lost in the measurement. It was shown that the phase can be retrieved in some cases by iterative algorithms using \textit{a priori} information about the object such as its size and symmetry. We suggest a more generalized and robust approach for the reconstruction of molecular orbitals based on state-of-the-art phase-retrieval algorithms currently used in coherent diffraction imaging. We draw an analogy between the phase problem in molecular orbital imaging by ARPES and of that in optical coherent diffraction imaging by performing an optical analogue experiment on micrometer-sized structures. We successfully reconstruct amplitude and phase of both the micrometer-sized objects and a molecular orbital from the optical and photoelectron far-field intensity distributions, respectively, without any prior information about the shape of the objects.
\end{abstract}

%
\vspace{2pc}

\section{Introduction}

Organic semiconductors play a key role in modern devices such as organic light-emitting diodes and photovoltaic cells~\cite{oregan1991, mathew2014}. More recently, organic molecules have been used as catalysts in photolytic water splitting, a promising route towards production of hydrogen as renewable energy source~\cite{li2015}. Tailoring the physical properties of molecular optoelectronic devices ~\cite{browne2009, pan2015, sharifzadeh2015} crucially depends on a deep understanding of the charge transfer mechanisms at metal-organic interfaces. The time-resolved spatial visualization of such processes would hence be highly desirable.

The frontier orbitals, i. e. the highest occupied (HOMO) and lowest unoccupied (LUMO) molecular orbitals, largely determine the chemical reactivity and electronic properties of molecular systems. Detailed information about the electronic structure of molecular systems can be inferred from angle-resolved photoelectron spectroscopy (ARPES) of well-ordered molecular layers on single-crystalline substrates~\cite{puschnig2009, puschnig2013, dauth2014, lueftner2014, wiessner2014, weiss2015}. The photoemission intensity

$I(\textit{\textbf{k}}_{\mathrm{f}\parallel}, E_{\mathrm{kin}})$ is derived from Fermi's golden rule as
\begin{eqnarray}
 I(\textit{\textbf{k}}_{\mathrm{f}\parallel}, E_{\mathrm{kin}})&\propto&\sum\limits_{i} \left| \right\langle\psi_\mathrm{f}(\textit{\textbf{k}}_{\mathrm{f}\parallel}, E_{\mathrm{kin}}, \textit{\textbf{r}}) \left| \textit{\textbf{A}}\cdot\textit{\textbf{p}}\left| \psi_i(\textit{\textbf{k}}_{i\parallel}, \textit{\textbf{r}})\right\rangle\right|^2 
 \nonumber \\
~&~&\times \delta(E_{\mathrm{kin}}+\Phi+E_{i}-\hbar\omega)\times\delta(\textit{\textbf{k}}_{\mathrm{f}\parallel}-\textit{\textbf{k}}_{i\parallel}-\textbf{G}_{\parallel}),
\end{eqnarray}
where $\psi_{i}$ and $\psi_{\mathrm{f}}$ denote initial and final state wave functions with corresponding momentum components 
$\textit{\textbf{k}}_{i\parallel}$ and $\textit{\textbf{k}}_{\mathrm{f}\parallel}$ parallel to the surface, respectively. The delta functions in the second line comprising photon energy $\hbar\omega$, sample work function $\Phi$ and reciprocal lattice vector $\textit{\textbf{G}}_{\parallel}$ ensure energy and momentum conservation in the photoemission process. The transition matrix element is given in the dipole approximation, where $\textit{\textbf{p}}$ and $\textit{\textbf{A}}$ denote the momentum operator and the vector potential of the exciting light. The photocurrent $I(\textit{\textbf{k}}_{\mathrm{f}\parallel}, E_{\mathrm{kin}})$ is obtained by summation over all transitions from occupied initial states $\psi_i$ to the final state $\psi_{\mathrm{f}}$ characterised by the kinetic energy $E_{\mathrm{kin}}$ and the parallel component of the final state momentum $\textit{\textbf{k}}_{\mathrm{f}\parallel}$ of the photoelectron. The photoemission final state $\psi_{\mathrm{f}}$ can be approximated by a plane wave $\propto \rme^{i\textit{\textbf{k}}_{\mathrm{f}}\textit{\textbf{r}}} $ provided the following conditions are fulfilled~\cite{puschnig2009, puschnig2013, goldberg1978}: (i) photoelectrons are emitted from $\pi$-orbitals of large planar molecules, for which all the contributing orbitals are of the same $p_z$ character; (ii) the molecules consist of mainly light atoms (H, C, N, O) and final state scattering effects can thus be neglected. Under these assumptions, the measured ARPES intensity becomes proportional to the squared modulus of the Fourier transform of the initial state wave function $\psi_i$ weakly modulated by a slowly varying angle-dependent envelope function~\cite{puschnig2009, puschnig2013}:
\begin{equation}
I(\textit{\textbf{k}}_{\mathrm{f}\parallel}, E_{\mathrm{kin}})\propto|\textit{\textbf{A}}\cdot\textit{\textbf{p}}|^2|\mathcal{F}\{\psi_i (\textit{\textbf{k}}_{i\parallel}, \textit{\textbf{r}})\}|^2
\end{equation}

The recorded intensity pattern, however, does not contain any information about the phase of the complex-valued electron wave distribution in the detector plane, which inhibits the direct reconstruction of the molecular wave function via computation of an inverse Fourier transform. In certain cases, phase information can be inferred from the parity of the wave function~\cite{puschnig2009} or from dichroism measurements~\cite{wiessner2014} and be imposed onto the measured data. However, the reconstruction of the molecular wave functions in such a way is not applicable to the most general type of problems when the phase distribution cannot be deduced from symmetry considerations. This issue was addressed by L\"uftner et al.~\cite{lueftner2014} by suggesting an iterative phase retrieval procedure similar to the Fienup algorithm~\cite{fienup1978}. In the suggested procedure, one iterates back and forth between real and reciprocal spaces by computing Fourier transforms and satisfying the constraints in both domains. In real space, the wave function is confined to a rectangular box which roughly corresponds to the van der Waals size of the molecule and thus represents the support of the object. The absolute value of the wave function is reduced to 10$\%$ outside this confinement box at each iteration step. In reciprocal space, the computed value of the amplitude is replaced by the measured one and the phase is kept.

In this work, we suggest that the phase problem in ARPES-based molecular orbital imaging can be solved in a more robust manner by utilizing the analogy to the phase problem in coherent diffraction imaging (CDI)~\cite{miao1999}. Both in CDI and orbital imaging, the far field pattern in the detector plane is proportional to the squared modulus of the Fourier transform of the object distribution. Provided the far-field intensity pattern is measured at the oversampling condition~\cite{miao1998, miao2003}, both the amplitude and the phase of the object can be reconstructed from the experimentally available modulus of its Fourier transform using the phase retrieval algorithms~\cite{fienup1978} as it is done in CDI~\cite{miao1999}. Therefore, we suggest to directly apply state-of-the-art phase-retrieval algorithms, currently used in CDI, for the reconstruction of molecular orbitals. These algorithms were specifically optimized for objects described by a complex-valued transmission function~\cite{harder2010}, which makes them ideal for the reconstruction of electron wave functions. Moreover, recent advancements in CDI allowed for the solution of the phase problem without need for the precise knowledge of the shape of the object, which is instead found in the course of the reconstruction using the shrinkwrap algorithm~\cite{marchesini2003}. To facilitate a better understanding of the CDI phase-retrieval algorithms in view of their applicability to reconstruction of molecular wave functions, we designed an optical analogue experiment and performed CDI on micrometer-sized structures produced by means of photolithography. Available CDI phase-retrieval algorithms~\cite{fienup1978, harder2010, marchesini2003} were employed for the reconstruction of the micrometer-sized object. Eventually, the same algorithms were applied to a set of ARPES data and the lowest unoccupied molecular orbital (LUMO) of pentacene was reconstructed.

\section{Methods}
\subsection{Optical CDI of a microstructure}

The microstructures for the optical CDI experiments were patterned in a 105 nm-thick Cr film deposited on a 1.7 mm-thick fused silica substrate, thus providing transparent objects in a non-transparent medium. The individual microstructures had an identical shape but different sizes and were separated from one another by several millimeters to avoid interference between the neighbouring objects. The size of the microstructures was selected in such way that the ratio between microstructure length (e.g.,  $\SI{15}{\micro m}$) and employed laser wavelength ($\SI{0.532}{\micro m}$) was comparable to the ratio between length of pentacene molecule ($\approx \SI{1.5}{nm}$) and de Broglie wavelength of the electrons ($\approx \SI{0.17}{nm}$) at the used photon energy ($\SI{50}{eV}$). The experimental setup for optical CDI is shown in Fig.~1. The laser beam profile had a Gaussian distribution as shown in the inset. For CDI experiments, the laser beam is usually spatially filtered and then expanded using two lenses, which ensures that the intensity profile in the object plane is constant~\cite{thibault2007}. In our experiment, we employed the laser beam without expansion because the light intensity variations on the length-scale of the microstructure were negligible. The microstructure was illuminated from the side of the Cr film. The far-field distribution of the scattered wave was imaged onto a semitransparent screen and the diffraction patterns were recorded with a 10-bit CCD camera (Hamamatsu C4742-95) placed behind the screen as shown in Fig.~1. In order to increase the dynamic range, we recorded several diffraction patterns at different exposures by using a rotatable neutral density filter with optical densities ranging from 0 to 4.0. The recorded images were then combined into one high-dynamic-range (HDR) image by a procedure proposed by Debevec~\cite{devebec1997}. 

 \begin{figure}[h]
 \includegraphics{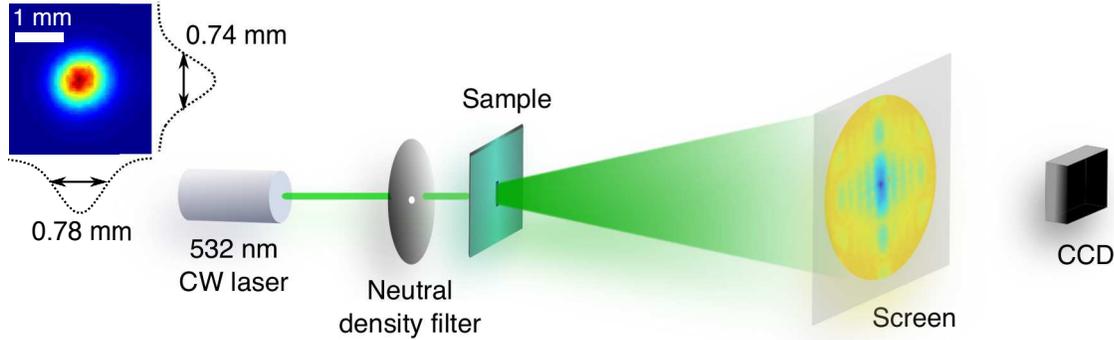}
 \caption{\label{fig1} Experimental setup of optical CDI. The distance between sample and screen was set to 22.5 cm. The size of the imaged screen area comprised $40\times40$ cm$^2$ sampled with $1000\times1000$ pixels. Inset: Intensity distribution of the laser profile.}
 \end{figure}
 
\subsection{ARPES of pentacene/Ag(110)}
A well-ordered sub-monolayer of pentacene molecules adsorbed on Ag(110) served as model system for  orbital tomography. Pentacene ARPES data has been acquired during a beamtime of A. Sch\"oll and coworkers  (University of W\"urzburg) at the NanoESCA beamline at Elettra synchrotron (Trieste, Italy) and has been provided to us for validation of our phase retrieval algorithm~\cite{grimm}. The crystal was prepared according to standard procedures~\cite{feyer2014} and pentacene molecules~\cite{puschnig2009} were deposited from a home-built Knudsen cell~\cite{wiessner2014}. ARPES constant binding energy (CBE) momentum maps of the pentacene LUMO were recorded with the p-polarized light at a photon energy of $\SI{50}{eV}$ using the photoemission electron microscope (PEEM)~\cite{schneider2012, patt2014}. The setup of the PEEM and the experimental geometry are shown in Fig.~2(a) and Fig.~2(b), respectively. The microscope was operated in the momentum mode and allowed for detection of electrons with the acceptance angle of $\alpha=\pm90^0$ corresponding to slightly less than $\pm \SI{3}{\AA^{-1}}$ at $\SI{50}{eV}$ photon energy without any sample rotation. The CBE map was integrated over a $\SI{200}{meV}$ energy window, which is of the order of the electron analyzer resolution and of the full-width at half-maximum of the pentacene LUMO at the binding energy of $\SI{0.1}{eV}$.
\begin{figure}[h]
 \includegraphics{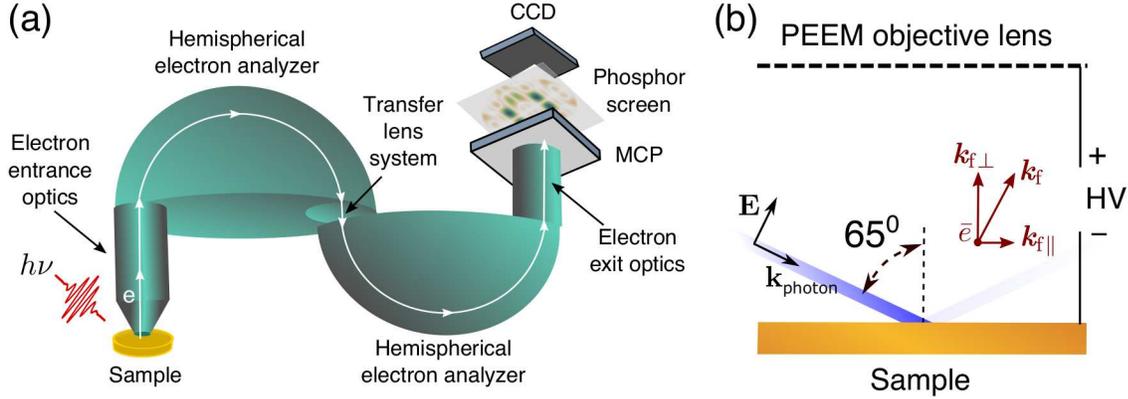}
\caption{\label{fig2} (a) Schematic of the PEEM setup. (b) Experimental geometry. The photon energy was 50 eV and the light was p-polarized with an incidence angle of  $65^0$. The photoemitted electrons were collected by the PEEM objective lens with an acceptance angle of $\alpha=\pm90^0$. $\textit{\textbf{k}}_{\mathrm{f}\parallel}$ and $\textit{\textbf{k}}_{\mathrm{f}\bot}$ denote parallel and normal components of the final state momentum of the photoelectrons.}
\end{figure}

\subsection{Algorithms}
Prior to reconstruction of the pentacene LUMO, we tested the performance of the algorithms on the optical CDI data set, taking advantage of the high dynamic range of these data. We employed a combination of the phase-constrained~\cite{harder2010} hybrid input-output~\cite{fienup1978} (PC-HIO) and error reduction~\cite{fienup1978} (ER) algorithms. The usage of both algorithms in an alternating scheme has been shown to eliminate stagnation problems and to provide faster convergence~\cite{fienup1978, harder2010, williams2006}.
 \begin{figure}[h]
 \includegraphics{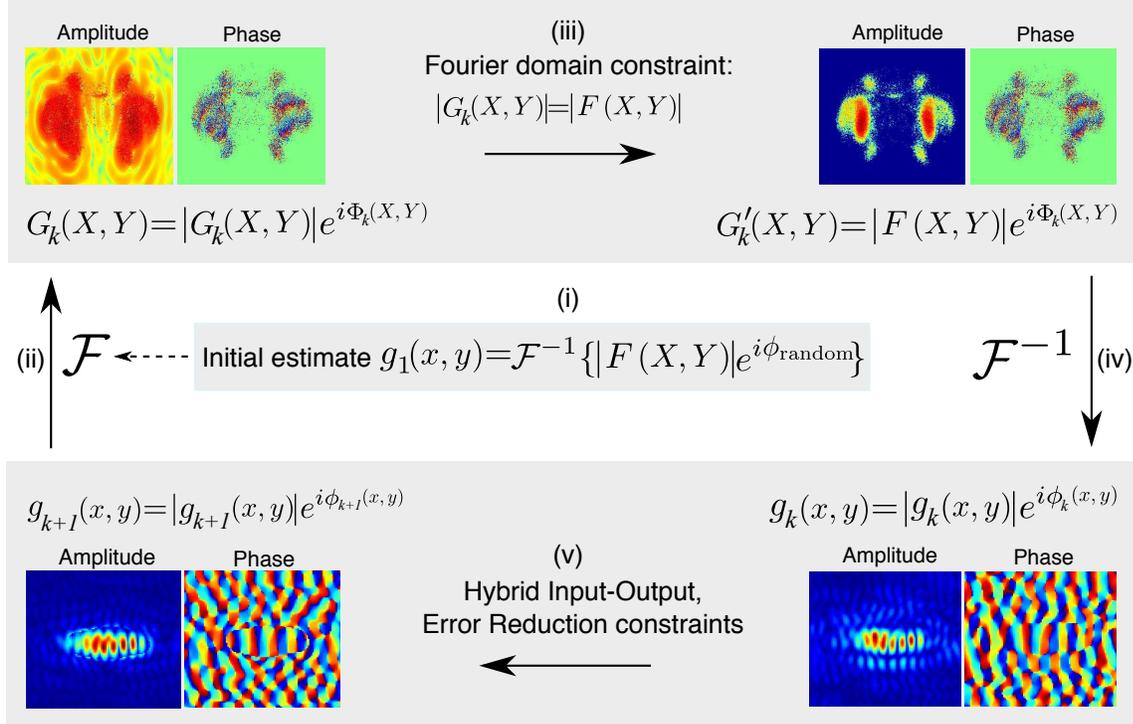}
 \caption{\label{fig3} Iterative phase retrieval scheme.}
 \end{figure}
The support of the object was found using the shrinkwrap algorithm~\cite{marchesini2003}. Following the conventional procedure of this algorithm, the initial estimate of the object support was obtained from the autocorrelation of the object by computing the inverse Fourier transform of the experimental diffraction pattern $I(X,Y)$, convolving it with a Gaussian function (width $\sigma=$ 5 pixels) and applying a threshold at $10\%$ of its maximum. The pixel values below the threshold were zeroed. The reconstruction began with 40 iterations of the PC-HIO algorithm followed by 2 iterations of the ER algorithm. We found that this number of iterations is sufficient to yield a resonable estimate of the object shape and thus to perform the first update of the object support by using the shrinkwrap procedure~\cite{marchesini2003} described in detail below. The scheme of the iterative phase retrieval procedure is shown in Fig.~3, which included the following steps:
\begin{enumerate}

\item In the first iteration $k=1$, the experimental amplitude $|F(X,Y)|=\sqrt{I(X,Y)}$ was combined with a random phase and the inverse Fourier transform supplied an initial input object distribution $g_k(x,y)$, where $(X,Y)$ and $(x,y)$ denote the coordinates in the detector and object planes, respectively. We assume the most general case of a complex-valued object distribution and keep both its real and imaginary parts.

\item By computing the Fourier transform of $g_k(x,y)$, we obtain the complex-valued distribution $G_k(X,Y)=\mathcal{F} \left\{g_k(x,y)\right\}$.

\item By replacing the calculated amplitude $|G_k(X,Y)|$ with the experimental amplitude $|F(X,Y)|$, while keeping the calculated phase distribution, we obtain an updated complex-valued field distribution in the detector plane $G'_k(X,Y)$. 

\item Inverse Fourier transform of $G'_k(X,Y)$ provides the output object distribution $g'_k(x,y)$. 

\item In the PC-HIO algorithm ~\cite{fienup1978, harder2010}, the input object for the next iteration $g_{k+1}(x,y)$ is obtained as

\begin{equation}
  g_{k+1}(x,y)=\left\{
  \begin{array}{@{}ll@{}}
    g'_k(x,y), & \text{if}\ (x,y)\in \gamma, \\
    g_k(x,y)-\beta g'_k(x,y), & \text{if}\ (x,y)\notin \gamma,
  \end{array}\right.
\end{equation} 
where $\beta=0.9$ is a feedback parameter and $\gamma$ corresponds to a set of points which comply with the object domain constraints (belong to the support region and have their phases within an expected range). In the ER algorithm~\cite{fienup1978}, the object distribution $g_{k+1}(x,y)$ is calculated as 

\begin{equation}
  g_{k+1}(x,y)=\left\{
  \begin{array}{@{}ll@{}}
    g'_k(x,y), & \text{if}\ (x,y)\in \gamma, \\
    0, & \text{if}\ (x,y)\notin \gamma,
  \end{array}\right.
\end{equation} 
where $\gamma$ fulfills the same criteria as in the PC-HIO algorithm. 

\end{enumerate}

The output object distribution $g'_k(x,y)$ obtained in the last iteration of the ER cycle was used to update the object support. This was done by convolving $g'_k(x,y)$ with a Gaussian function and setting a threshold at $12\%$ of its maximum, as it is typically done in the shrinkwrap algorithm~\cite{marchesini2003}. The width of the Gaussian was initially set to 2.5 pixels. Upon the first update of the support, the algorithm continued with alternating cycles of 20 iterations of the PC-HIO algorithm followed by 2 iterations of the ER algorithm~\cite{fienup1978, harder2010, williams2006}. The end of each cycle was finalized by computing a new distribution of the object support. The threshold value and the Gaussian width were chosen empirically so that no part of the reconstructed pattern was truncated, but instead the support converged smoothly towards the shape of the object. The latter requirement was ensured by reducing the width of the Gaussian at every support update by $1\%$ as it is conventionally done in the shrinkwrap algorithm~\cite{marchesini2003}. The quality of the reconstructions was estimated by computing the mismatch between the iterated and the experimental amplitudes~\cite{fienup1978, fienup1982, fienup1986}:  

\begin{equation}\label{eq:error}
E=\sqrt{\frac{\sum_{X,Y=0}^{N-1}||F(X,Y)|-|G_{\mathrm{it}}(X,Y)||^2}{\sum_{X,Y=0}^{N-1}|F(X,Y)|^2}},
\end{equation} 
where $|F(X,Y)|$ is the experimental amplitude, $|G_{\mathrm{it}}(X,Y)|$ is the iteratively obtained amplitude.

\subsection{Oversampling requirements}

The solution of the phase problem requires the fulfillment of the oversampling condition~\cite{miao1998, miao2003}. Given an $N\times N$ pixel sampled amplitude $|F(X,Y)|=|\sum_{X,Y=0}^{N-1} f(x,y) e^{-2\pi i(xX+yY)/N}|$ in reciprocal space, we obtain a set of $N^2$ equations, which have to be solved in order to find both the amplitude and phase of $f(x,y)$. Miao et al.~\cite{miao1998} defined the oversampling ratio as

\begin{equation}
\sigma=\frac{N_{\mathrm{total}}}{N_{\mathrm{unknown}}},
\end{equation}

where $N_{\mathrm{total}}$ is the total number of pixels and $N_{\mathrm{unknown}}$ is the number of pixels with unknown values. The set of equations is solved by dense sampling of the diffraction pattern so that the object distribution is surrounded by a zero-padded region with $\sigma>2$~\cite{miao1998}. In each dimension of a 2D data set, we can define a linear oversampling ratio

\begin{equation}\label{eq:los}
\O=\frac{N\Delta r}{a},
\end{equation}

where $N$ is the linear number of pixels, $\Delta r$ is the size of the pixel in the object domain and $a$ is the largest extent of the object. The oversampling requirement then corresponds to  $\O>\sqrt{2}$~\cite{miao1998}.

\section{Results and discussion}

\subsection{Optical CDI: Reconstruction of the micrometer-sized structures}

Fig.~4 shows the results of the reconstruction of the micrometer-sized structures. In optical CDI, we employed micrometer-sized structures of $30\times12$ $\SI{}{\micro m^2}$ (sample 1) and $14.8\times6$ $\SI{}{\micro m^2}$ (sample 2). The scanning electron microscope (SEM) images of samples 1 and  2 are shown in Fig.~4(a,b) next to the experimental diffraction patterns (Fig.~4(c,d)). The size of the diffraction patterns sampled with $1000\times1000$ pixel was $40\times40$ cm$^2$ in each case, thus giving the size of the pixel in the detector plane $\Delta p=400$ $\SI{}{\micro m}$. The size of the pixel in the object plane $\Delta r$ can be related to the distance $z=22.5$ cm from the object to the detector plane and to the employed laser wavelength $\lambda=532$ nm~\cite{zuerch2015}. The linear oversampling ratio defined by Eq.~\ref{eq:los} can be rewritten as~\cite{miao1998}:

\begin{equation}
\O=\frac{z\lambda}{a\Delta p}
\end{equation}

For the samples 1 and 2 with the lengths $a_1=\SI{30}{\micro m}$ and $a_2=\SI{15}{\micro m}$, the linear oversampling ratios fulfilled the oversampling condition and were $\O_1\approx10$ and $\O_2\approx20.2$, respectively.

 \begin{figure}[h]
 \includegraphics{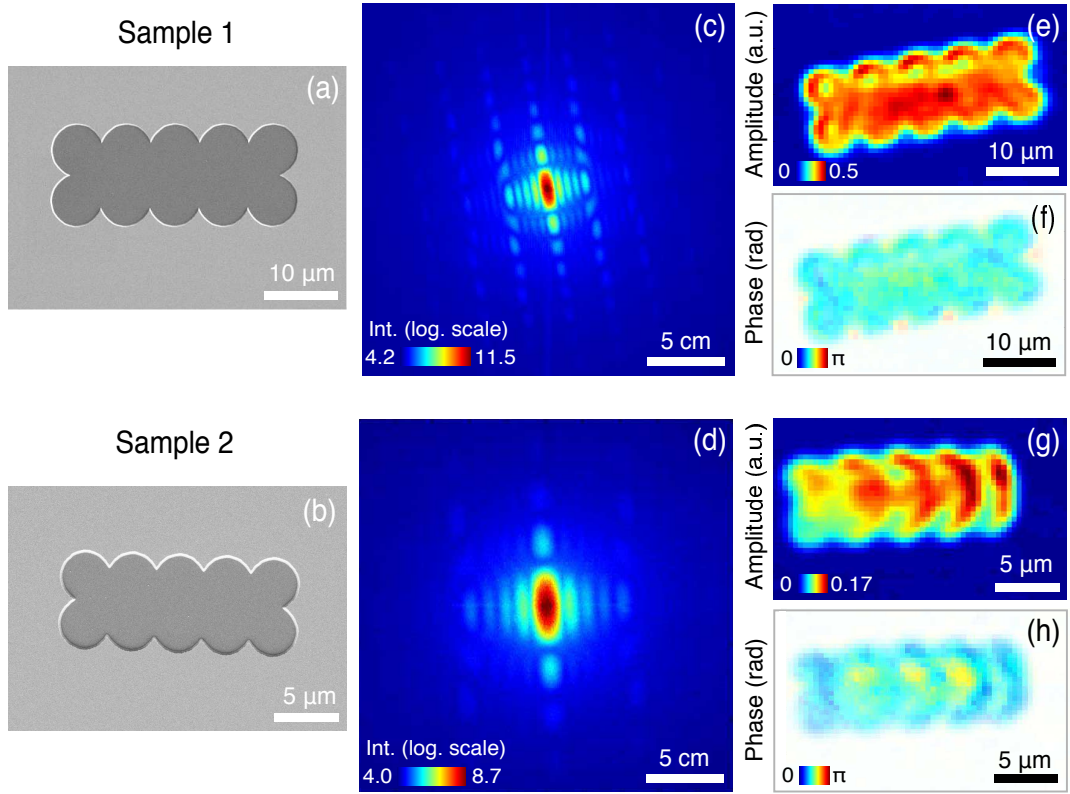}
 \caption{\label{fig4} Reconstruction of the micrometer-sized objects. (a,b) SEM images. (c,d) Experimental diffraction pattern intensities shown on logarithmic scale. (e,g) Reconstructed amplitudes. (f,h) Reconstructed phases.}
 \end{figure}
Prior to application of the phase retrieval algorithms, the experimental diffraction patterns were pre-processed: First, each of the recorded $1000\times1000$ pixel images was centered. Centering of the experimental diffraction pattern was shown to have a strong effect on the quality of the reconstruction in CDI~\cite{zuerch2013}. The noise of the CCD camera (average count rate of 50 counts) was subtracted from each pixel and the images were truncated to $500\times500$ pixels around their centers because of the low signal-to-noise ratio at the peripheral parts. The central part of each diffraction pattern was dominated by an intense laser spot due to the partial transparency of the chromium film to the laser beam. Pixel values exceeding the thresholds of $1.5\cdot10^5$ counts (sample 1) and $6\cdot10^3$ counts (sample 2) were defined as missing and their values were updated in the course of the reconstruction by using the corresponding pixel values of the calculated amplitudes in the detector plane \cite{marchesini2003, latychevskaia2015}. In each case, the square root of the resulting diffraction pattern was fed into the algorithm. We found that 10 alternating cycles of the PC-HIO and ER algorithms, each followed by an update of the support, were enough to achieve a stable reconstruction. Further increase in the number of the reconstruction cycles was not necessary since it did not improve the quality of the reconstructed object distribution. At the end of 10 cycles, each reconstruction was stabilized by 100 iterations of the ER algorithm~\cite{latychevskaia2015}. In total, we performed 1000 independent reconstructions by employing a random phase distribution for each reconstruction run. Eventually, the 50 reconstructions with the smallest error $E$ as defined by Eq.~\ref{eq:error} were selected and averaged~\cite{latychevskaia2015} and are shown in Fig.~4(e-h). The reconstructed amplitudes correctly reproduce the shape and dimension of the microstructures. Furthermore, as it was expected for a purely transmitting object illuminated by a Gaussian beam with an almost planar wavefront at the object site, the phase distributions turned out to be almost constant. The lower quality of the reconstructed amplitude of sample 2 (Fig.~4 (g)) can be attributed to the low signal-to-noise ratio in the respective diffraction pattern.
 
\subsection{ARPES orbital tomography: Reconstruction of the pentacene LUMO} 
We then applied the same algorithm to the ARPES data. Fig.~5 shows the results of the reconstruction of the pentacene LUMO. The experimental CBE map is shown in Fig.~5(a). Given the resolution in reciprocal space of $\Delta k\approx$ $\SI{0.01}{\AA^{-1}}$ and the length of the pentacene molecule $a\approx\SI{15}{\AA}$, the linear oversampling ratio in the ARPES experiment can be calculated using Eq.~\ref{eq:los}. Taking the relation $\Delta r \Delta k = \frac{2\pi}{N}$ between the pixel size in object space, $\Delta r$, and reciprocal space, $\Delta k$, into account, the linear oversampling ratio can be expressed as

\begin{equation}
\O=\frac{2\pi}{a\Delta k}.
\end{equation}

The linear oversampling ratio was $\O\approx42$ and thus fulfilled the oversampling condition~\cite{miao1998}. The experimental CBE map was pre-processed following similar steps as those applied to the reconstruction of the micrometer-sized objects: First, the image was centered and the quasi-constant noise of the CCD camera (average count rate of 50 counts) was subtracted from each pixel. To ensure a sufficient number of pixels allocated per unit length of the molecule, we zero-padded the experimental CBE map to $2000\times2000$ pixels around its center. The square root of the processed CBE map was fed into the algorithm with the same parameters as used for the reconstruction of the micrometer-sized structures. Varying these parameters did not lead to any substantial improvements in the quality of the reconstruction. In total, we performed 1000 reconstructions of the pentacene LUMO. About $56\%$ of the reconstructed objects $g(x,y)$ were reconstructed together with their conjugate $g^*(-x,-y)$ or twin images~\cite{fienup1986}. The identification of the twin images could be automated by a procedure proposed by Fienup~\cite{fienup1986}, but here they were easily identified by visual inspection and discarded. From the remaining reconstructions, 50 with the smallest error $E$ as defined by Eq.~\ref{eq:error} were selected and averaged. The reconstructed amplitude and phase of the pentacene LUMO are shown in Fig.~5 (b-c) together with the overlayed carbon frame of the molecule for comparison. 
 \begin{figure}[h]
 \includegraphics{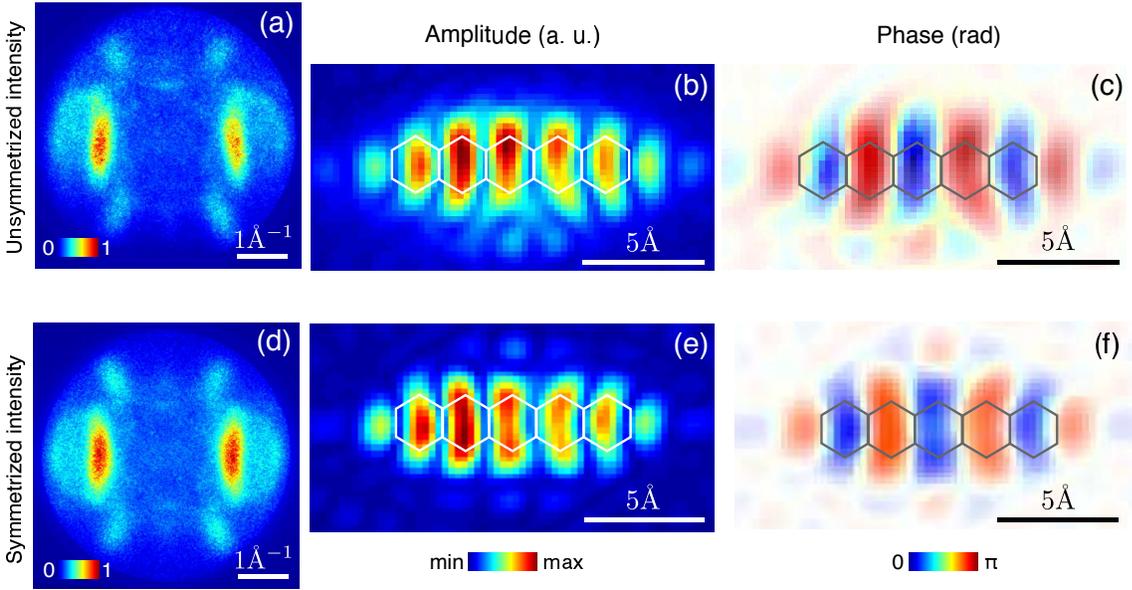}
 \caption{\label{fig5} Reconstruction of the pentacene LUMO. (a) CBE map recorded with PEEM from a sub-monolayer of pentacene on Ag(110) at 50 eV photon energy. (b) Reconstructed amplitude of the LUMO. (c) Reconstructed phase. Image transparency is weighted with the corresponding amplitude values for illustration purposes. (d) The same CBE map as in (a), but symmetrized with respect to the center. Reconstructions of (e) amplitude and (f) phase obtained from (d).}
 \end{figure}
 
It should be noted that we did not perform a normalization of the ARPES intensity by the angle-dependent factor $|\textbf{A}\cdot\textbf{p}|^2$ nor did we enforce any symmetry constraints in the course of the reconstruction onto the amplitude and phase shown in Fig.~5 (b-c). The object distribution was let to freely evolve until the stable solution was reached, which makes the utilized algorithm independent of any symmetry properties imposed onto the object under reconstruction. Furthermore, we note that the recorded CBE map shown in Fig.~5(a) contains features coming from the Ag(110) substrate (mostly at high momenta), but they do not seem to have a profound effect on the results of the reconstruction. By comparing our results with the literature, we find that the phase distribution weighed with the correspondent amplitude values as well as the shape of the orbital correctly reproduce the DFT calculations~\cite{lueftner2014, ules2014} as well as the data reconstructed by L\"uftner et. al~\cite{lueftner2014}. 

Finally, in order to assess the robustness of the algorithm in terms of the quality of reconstruction from the unsymmetrized CBE map, we made use of the symmetry properties of the pentacene LUMO amplitude and phase and symmetrized the CBE map shown in Fig.~5 (a) around its center. The symmetrical version is shown in Fig.~5 (d). In optics, the far field diffraction pattern is symmetric only in two cases: either due to the real-valued nature of an object or in case of an even complex-valued object distribution with an even amplitude and an even phase. In the latter case, the Fourier transform of the even complex-valued function is an even function as well and the far field intensity distribution is therefore symmetric. In the case of the complex-valued wave function of the pentacene LUMO, the symmetrization is justified purely due to the symmetry of the LUMO amplitude and the phase as it is known from the DFT calculations ~\cite{lueftner2014, ules2014}. The symmetrized CBE map was pre-processed following the same procedure as described above and the results of the reconstruction are shown in Fig.~5(d-f). Qualitatively, the reconstructions from the unsymmetrized CBE map are as good as the reconstructions from the symmetrized data set, except for some minor differences in the shapes of the lobes due to the intrinsic asymmetry of the CBE in Fig.~5(a). This agreement further proves the robustness of the employed algorithm for the reconstruction of molecular orbitals with arbitrary symmetry properties.

\section{Summary and Conclusion}

In this work, we show that the state-of-the-art phase retrieval algorithms currently employed in CDI can be successfully used for the reconstruction of complex-valued wave functions of molecules adsorbed on single-crystalline substrates. We tested and applied these algorithms in an optical analogue experiment and then successfully applied them to the reconstruction of the LUMO of pentacene adsorbed on Ag(110). The advantage of using modern CDI algorithms and in particular the shrinkwrap algorithm for the reconstruction of molecular orbitals is that they do not require any \textit{a priori} information about the shape of the object. Instead, they smoothly converge to the correct shape of the object in the course of the reconstruction. In case of molecular wave functions, this is highly important, since precise estimation of the object support is difficult and cannot be guaranteed in every case. This applies, for instance, if the orbital tomography technique aims at visualizing chemical reactions or following the dynamics of excited states, where effective electronic wave functions are unknown. The availability of a general and robust reconstruction algorithm is thus an important step for further advancement of orbital tomography. 

\ack
Financial support by the Swiss National Science Foundation through NCCR MUST is greatefully acknowledged. We thank Achim Sch\"oll and co-workers (University of W\"urzburg) for making the pentacene ARPES data available to us. The Center for Micro- and Nanoscience (ETH Zurich) is acknowledged for design and production of the photomask.  SEM imaging was performed with equipment maintained by the Center for Microscopy and Image Analysis (University of Zurich).

\section*{References}
\bibliography{iopart-num}

\providecommand{\newblock}{}
\begin{thebibliography}{10}
\expandafter\ifx\csname url\endcsname\relax
  \def\url#1{{\tt #1}}\fi
\expandafter\ifx\csname urlprefix\endcsname\relax\def\urlprefix{URL }\fi
\providecommand{\eprint}[2][]{\url{#2}}

\bibitem{oregan1991}
O'Regan B and Graetzel M 1991 {\em Nature\/} {\bf 353} 737--40

\bibitem{mathew2014}
Mathew S, Yella A, Gao P, Humphry-Baker R, Curchod B~F, Ashari-Astani N,
  Tavernelli I, Rothlisberger U, Nazeeruddin M~K and Graetzel M 2014 {\em Nat
  Chem\/} {\bf 6} 242--7

\bibitem{li2015}
Li F, Fan K, Xu B, Gabrielsson E, Daniel Q, Li L and Sun L 2015 {\em Journal of
  the American Chemical Society\/} {\bf 137} 9153--9

\bibitem{browne2009}
Browne W~R and Feringa B~L 2009 {\em Annual Review of Physical Chemistry\/}
  {\bf 60} 407--28

\bibitem{pan2015}
Pan H, Zhang X, Yang Y, Shao Z, Deng W, Ding K, Zhang Y and Jie J 2015 {\em
  Nanotechnology\/} {\bf 26} 295302

\bibitem{sharifzadeh2015}
Sharifzadeh S, Wong C~Y, Wu H, Cotts B~L, Kronik L, Ginsberg N~S and Neaton J~B
  2015 {\em Advanced Functional Materials\/} {\bf 25} 2038--46

\bibitem{puschnig2009}
Puschnig P, Berkebile S, Fleming A~J, Koller G, Emtsev K, Seyller T, Riley J~D,
  Ambrosch-Draxl C, Netzer F~P and Ramsey M~G 2009 {\em Science\/} {\bf 326}
  702--6

\bibitem{puschnig2013}
Puschnig P, Koller G, Draxl C and Ramsey M~G The structure of molecular
  orbitals investigated by angle-resolved photoemission {\em Small Organic
  Molecules on Surfaces\/} ed Sitter H, Draxl C and Ramsey M

\bibitem{dauth2014}
Dauth M, Wiessner M, Feyer V, Schoell A, Puschnig P, Reinert F and Kuemmel S
  2014 {\em New Journal of Physics\/} {\bf 16} 103005

\bibitem{lueftner2014}
Lueftner D, Ules T, Reinisch E~M, Koller G, Soubatch S, Tautz F~S, Ramsey M~G
  and Puschnig P 2014 {\em PNAS\/} {\bf 111} 605--10

\bibitem{wiessner2014}
Wiessner M, Hauschild D, Sauer C, Feyer V, Schoell A and Reinert F 2014 {\em
  Nature Communications\/} {\bf 5} 4156

\bibitem{weiss2015}
Weiss S, Lueftner D, Ules T, Reinisch E~M, Kaser H, Gottwald A, Richter M,
  Soubatch S, Koller G, Ramsey M~G, Tautz F~S and Puschnig P 2015 {\em Nature
  Communications\/} {\bf 6} 8287

\bibitem{goldberg1978}
Goldberg S~M, Fadley C~S and Kono S 1978 {\em Solid State Communications\/}
  {\bf 28} 459--463

\bibitem{fienup1978}
Fienup J~R 1978 {\em Optics Letters\/} {\bf 3} 27--9

\bibitem{miao1999}
Miao J, Charalambous P, Kirz J and Sayre D 1999 {\em Nature\/} {\bf 400} 342--4

\bibitem{miao1998}
Miao J, Sayre D and Chapman H~N 1998 {\em J. Opt. Soc. Am. A\/} {\bf 15}
  1662--9

\bibitem{miao2003}
Miao J, Ishikawa T, Anderson E~H and Hodgson K~O 2003 {\em Physical Review B\/}
  {\bf 67} 174104

\bibitem{harder2010}
Harder R, Liang M, Sun Y, Xia Y and Robinson I~K 2010 {\em New Journal of
  Physics\/} {\bf 12} 035019

\bibitem{marchesini2003}
Marchesini S, He H, Chapman H~N, Hau-Riege S~P, Noy A, Howells M~R, Weierstall
  U and Spence J~C~H 2003 {\em Physical Review B\/} {\bf 68} 140101

\bibitem{thibault2007}
Thibault P and Rankenburg I~C 2007 {\em American Journal of Physics\/} {\bf 75}
  827

\bibitem{devebec1997}
Devebec P~E and Malik J 1997 Recovering high dynamic range radiance maps from
  photographs {\em Proceedings of the 24th annual conference on Computer
  graphics and interactive techniques\/} (ACM Press/Addison-Wesley Publishing
  Co.)

\bibitem{grimm}
Grimm M, Graus M, Metzger C and Schoell A 2016 unpublished data

\bibitem{feyer2014}
Feyer V, Graus M, Nigge P, Wiessner M, Acres R~G, Wiemann C, Schneider C~M,
  Schoell A and Reinert F 2014 {\em Surface Science\/} {\bf 621} 64--8

\bibitem{schneider2012}
Schneider C, Wiemann C, Patt M, Feyer V, Plucinski L, Krug I, Escher M, Weber
  N, Merkel M, Renault O and Barrett N 2012 {\em Journal of Electron
  Spectroscopy and Related Phenomena\/} {\bf 185} 330

\bibitem{patt2014}
Patt M, Wiemann C, Weber N, Escher M, Gloskovskii A, Drube W, Merkel M and
  Schneider C~M 2014 {\em Review of Scientific Instruments\/} {\bf 85} 113704

\bibitem{williams2006}
Williams G~J, Pfeifer M~A, Vartanyants I~A and Robinson I~K 2006 {\em Physical
  Review B\/} {\bf 73} 094112

\bibitem{fienup1982}
Fienup J~R 1982 {\em Applied Optics\/} {\bf 21} 2758--69

\bibitem{fienup1986}
Fienup J~R and Wackerman C~C 1986 {\em Journal of the Optical Society of
  America\/} {\bf 3} 1897--1907

\bibitem{zuerch2015}
Zuerch M~W 2015 {\em High-Resolution Extreme Ultraviolet Microscopy\/} Springer
  Theses (Springer International Publishing)

\bibitem{zuerch2013}
Zuerch M, Kern C and Spielmann C 2013 {\em Optics Express\/} {\bf 21} 21131--47

\bibitem{latychevskaia2015}
Latychevskaia T, Chushkin Y, Zontone F and Fink H~W 2015 {\em Applied Physics
  Letters\/} {\bf 107} 183102

\bibitem{ules2014}
Ules T, Lueftner D, Reinisch E~M, Koller G, Puschnig P and Ramsey M~G 2014 {\em
  Physical Review B\/} {\bf 90} 155430

\end{thebibliography}

\end{document}